\newcommand{\defeq}[1]{\begin{equation}\label{#1}}
\newcommand{\eeq}{\end{equation}}
\documentclass{article}

\usepackage{amsfonts}
\usepackage{amsmath}

\usepackage{amssymb}
\usepackage{amsthm}

\usepackage{framed}
\usepackage{algorithm}
\usepackage{algorithmicx}
\usepackage{algpseudocode}

\usepackage[margin=1.0in]{geometry}
\usepackage{multicol}
\usepackage{multirow}
\usepackage{bigdelim}

\usepackage{etoolbox}

\usepackage{color} 
\definecolor{shadecolor}{gray}{0.9}
\definecolor{lightyellow}{RGB}{255,255,192}
\definecolor{lightblue}{RGB}{192,192,255}
\definecolor{lightpurple}{RGB}{255,192,255}
\definecolor{lightgreen}{RGB}{192,255,192}
\usepackage{mdframed}

\usepackage{array}
\usepackage{colortbl}
	\definecolor{CellColor}{RGB}{255,255,192}
	\definecolor{HighlightColor}{RGB}{255,192,192}
	\definecolor{HighlightColorB}{RGB}{192,255,192}

	\newcolumntype{C}{>{\columncolor[RGB]{255,255,192}}c}
	\newcolumntype{L}{>{\columncolor[RGB]{255,255,192}}l}
	\newcolumntype{R}{>{\columncolor[RGB]{255,255,192}}r}
	\arrayrulecolor[RGB]{0,0,0}
	\doublerulesepcolor[RGB]{0,0,0}
\usepackage{hhline}

\usepackage{float}
\usepackage{caption}[2007/09/01] 

\usepackage{tikz}
	\usetikzlibrary{arrows}
	\usetikzlibrary{decorations.fractals}
	\usetikzlibrary{decorations.pathreplacing}
	\usetikzlibrary{patterns}
	\usetikzlibrary{shapes}

\usepackage{outlines}
\usepackage{enumitem}
	\setenumerate[4]{label=\alph*.}
	\setenumerate[1]{label=\Roman*.}
	\setenumerate[2]{label=\Alph*.}
	\setenumerate[3]{label=\roman*.}

\usepackage{url}
\usepackage{hyperref}		

\numberwithin{equation}{section}	

\theoremstyle{plain}			

\theoremstyle{definition}		

\renewcommand{\qed}{\hfill \mbox{\raggedright \rule{.07in}{.1in}}}

\DeclareCaptionLabelFormat{underlcap}{\bf{$\underline{\text{Figure #2}}$}}
\newenvironment{PageFigure}
{
	\begin{table*}
	\begin{framed}
}
{
	\end{framed}
	\end{table*}
}

\newcommand\Section[2][]{\begin{mdframed}[skipabove=12pt,skipbelow=0.5cm, leftline=false, rightline=false, bottomline=true, topline=false, linecolor=black, linewidth=2,innertopmargin=10pt,innerbottommargin=10pt]   \ifx\relax#1\relax\section{#2}\else\section[#1]{#2}\fi   \end{mdframed}} 

\newcommand\Subsection[2][]{\begin{mdframed}[skipabove=12pt,skipbelow=0.5cm, leftline=false, rightline=false, bottomline=true, topline=false, linecolor=black, linewidth=2,innertopmargin=10pt,innerbottommargin=10pt]   \ifx\relax#1\relax\subsection{#2}\else\subsection[#1]{#2}\fi   \end{mdframed}} 

\patchcmd{\thebibliography}{\section*}{\Section}{}{}

\newcommand{\TitleString}{Faster Integer Multiplication Using Preprocessing} 
\title{\TitleString\\
\copyright \small{Matt Groff 2019} \\
} 
\author{MATT GROFF\\
P.O. Box 642\\
Camp Hill, PA, USA 17001-0642\\
mgroff100@hotmail.com}

\begin{document}
\pagestyle{myheadings}

\markright{\rlap{Multiplication Using Preprocessing}\hfill \textnormal{\copyright Matt Groff 2019}\hfill}


\twocolumn[\maketitle]


\begin{abstract}

A New Number Theoretic Transform(NTT), which is a form of FFT, is introduced, that is faster than FFTs.  Also, a multiplication algorithm is introduced that uses this to perform integer multiplication faster than $O(n \log{n})$.  It uses preprocessing to achieve an upper bounds of $(n \log{n} / (\log{\log{n}} / \log{\log{\log{n}}})$.

Also, we explore the possibility of $O(n)$ time multiplication via NTTs that require only $O(n)$ operations, using preprocessing.

\end{abstract}

\Section{Introduction}
Our algorithm, like many of the multiplication algorithms before, relies on the DFT, and that DFT is now the major bottleneck in multiplication algorithms due to its' $O(n \log{n})$ time.


The DFT is a very useful algorithm.  With the popularization of the FFT by Cooley and Tukey in 1965\cite{Cooley-Tukey-DFT}, it became much more widely used.  Since then, there have been a few attempts at speeding up the DFT.  In 2012 there was a major breakthrough with Hassanieh, Indyk, and Price's sparse FFT (sFFT)\cite{arxiv:sFFT}, that gave a time of $O(k \log{n})$ for exactly $k$-sparse algorithms and $O(k \log{n} \log{n/k})$ for the general case.  One drawback of the algorithm is that it needs $n$ to be a power of 2.

In this paper, we present a DFT algorithm that uses only in $O(n \log{n} / \log{\log{n}})$ operations, but it must use preprocessing.  Like some forms of the DFT, it only works for certain sizes, although it is much less limited than many.  Then we use this to multiply two naturals in the same time.  Essentially, we prove that integer multiplication can be done in time slightly faster than $O(n \log{n})$.

We note that it was Karatsuba, who in 1962, first improved on the na{\"{i}}ve bound of $n^2$ to ${n^{\log_2{3}}}$.

Besides the straight line program asymptotic limit \cite{Straight-Line-Bounds}, Sch{\"o}nhage and Strassen conjectured in 1971 that integer multiplication must take time $\Theta{(n \log{n})}$\cite{Schonhage-Strassen}.  In that same paper, they set the world record at $O(n \log{n} \log{\log{n}})$.  That bound stood for almost 40 years, until F{\"u}rer got the new bounds $O \left( n \log{n}K^{\log^*{n}} \right)$, where $K$ is some constant and $\log^*{}$ is the iterated logarithm\cite{Furer-paper}.  Finally, there were a number of results, particularly with a number of impressive papers from Harvey and van der Hoeven, that culminated in their $O(n \log{n})$ time algorithm\cite{Harvey-van-der-Hoeven}.  

It should be noted that in this paper, the focus is on the DFT, so we use $N$ for the size of the DFT, $\rho$ for the size of the multiplication, and we do calculations modulo a prime $P$.

\Section{Algorithm Overview\label{Overview Section}}
We will use the unit-cost RAM model for our algorithm.  We assume that we're multiplying two $n$ bit integers $A$ and $B$ together.  We use the convention that the maximum size of the input integers to the multiplication problem is $N$, which of course require $n$ bits.  Similarly, we use a prime of maximum size $P$, with $p = \log{P}$.  Similarly, we break up values of maximum size $N$ into ``digits'' of maximum size $R$, where $r = \log{R}$.  We use $R$ for splitting apart our calculations into smaller digits that we can use for quicker calculations, and mainly, for smaller lookup tables.  We also let $\nu = N \log{N} = Nn$, $\pi = P \log{P} = Pp$, and $\rho = R \log{R} = Rr$.  We use $x_j$ to denote the $j$th input to the DFT and $\widehat{x_k}$ to denote the $k$th output.


Later, we will use a multiplication algorithm where we assume that we are multiplying two $s$-bit numbers.

The algorithm gets its' speed advantage from the ability to solve small DFTs in linear number of operations, via table lookup.  The basic idea is then to use divide-and-conquer methodology to transform each size $n$ DFT into many DFTs of approximately size $\sqrt{n}$.  After enough recursion, the algorithm reaches DFTs that are sufficiently small enough.  The algorithm uses another table to convert each input coefficient into many coefficients of much smaller size, using the Chinese remainder theorem.  Then, the algorithm uses a table, as mentioned in the beginning, to lookup the result of this much smaller DFT.  It uses one more table to combine the smaller outputs into outputs of the original size, again via Chinese remainder theorem.  The recursion is then finished and the final result is output.  We note that the conversion to smaller size coefficients and table lookup is done only at the ``innermost'' level of the recursion, and is not done at every step of the recursion.

\Subsection{Recursion Primer}
This section borrows liberally from \cite{website:Wikipedia-Cooley-Tukey}.

We can write the DFT as the following formula:

\defeq{DFT Formula}
	\widehat{x_k} = \sum_{j=0}^{n-1}{ x_j \omega^{jk} }
\eeq

Here the $j$th input is $x_j$, and the $k$th output is $\widehat{x_k}$.  $\omega$ represents an $n$th \textit{principal} root of unity, which is necessary for the DFT to function correctly.

We can now rewrite $n = n_1 \cdot n_2$, and rewrite our function to accommodate recursion.  We set $j = j_1 n_1 + j_0$ and $k = k_2 n_2 + k_0$.  Then our original function becomes

\begin{align}
	\widehat{x_k}                 &= \sum_{j=0}^{n-1}{ x_j \omega^{jk} } \\
	                              &= \sum_{j_0 = 0}^{n_2-1}{ \sum_{j_1=0}^{n_2-1}{ x_{\underbrace{j_1 n_1 + j_0}_j} \omega^{\underbrace{(j_1 n_1 + j_0)}_j \underbrace{(k_2 n_2 + k_0)}_k } } } \\
																&= \sum_{j_0 = 0}^{n_2-1}{ \underbrace{\omega^{j_0 k_0}}_{\substack{\text{twiddle} \\ \text{factors}}} \underbrace{ \left( \sum_{j_1=0}^{n_2-1}{ x_{j_1 n_1 + j_0} \omega^{n_1 j_1 k_0}} \right) }_\text{Inner DFT} \omega^{n_2 j_0 k_2} }
\end{align}

This shows how we can express one DFT as many DFTs of two particular sizes, namely $n_1$ and $n_2$.  Splitting up one transform into more than one transform (size) is known as the \textit{multidimensional DFT}.

\Subsection{Tabular DFT \label{Tabular Introduction Section}}
Currently, for our DFTs, we perform calculations in a finite field, and we will say that we do most calculations modulo a prime $P$, with $P > n$, where $n = \log{N}$ is the total number of coefficients.  This is known as the \textit{number theoretic transform}.  

We will find that the size of the prime $P$ that we need to use will be approximately $O(n^L)$, where $n$ is the actual size of the DFT, and $L$ is a constant known as \textit{Linnik's constant}.  In order to use tables to find the DFT result more quickly, we want to reduce our calculations from modulo $P \approx n^L$ to modulo a much smaller value, say mod $r \approx n / 4$.  This will ensure that we can multiply two $r$ bit numbers together to get a number less than $2r$ bits, which means that we only need a table of $2^{2r}$ entries in it to record all possible multiplications of numbers of this size.  This will allow us to have a maximum of $n / (2r) = 2r$ bits per entry, and therefor the table won't be larger than $n$ total bits.  We can then call our temporary ``word'' size for this algorithm as $R = 2^r = 2^{n^{1/4}}$.

Now that we know the word size, the rest is more straightforward.  First, to multiply two values $a$ and $b$ together modulo $P$, we split them up into $a_j R^j$ and $b_k R^k$ and 

\begin{align}
	(a) \cdot (b) &= \\
	\left( \sum_j{ a_j R^j } \right) \left( \sum_k{ b_k R^k } \right) &= \
\label{Combined Equation}	\sum_{j+k = \ell}{ a_j b_k R^{\ell} }
\end{align}

We can simplify Equation \ref{Combined Equation} by focusing on calculating $a_j b_k R^{\ell}$.  If we use a table for each $\ell$ from $0$ to $2\cdot(4L) = 8L$, we can calculate any possible value we will need.  This is because there are at most $4L$ powers of $w$ in $a$ and $b$, and their product requires, at most, double this number.  Now this number is an exact value modulo $P$, which again is $a_j b_k R^{\ell}$.  We record exactly this value, modulo $P$, in our table.


Since we are adding together $8L$ numbers to get our result, we can reduce them to a value, modulo $P$, with $\log{8L} = 3+\log{L}$ comparisons.  This is because the final number, itself, will be at most $P \log{8L}$, due to adding $8L$ results from multiplication together.  So we compare this number to $P (8L)$, and if is is greater, then we subtract $P (8L)$.  Next, compare to $P (8L) / 2$.  If it is greater, subtract $P (8L) / 2$.  We repeat this $\log{8L}$ times, which is a constant.  Thus we can multiply two values less than $P$ in a constant number of arithmetic operations.

\Subsection{The ``Leaves'' \label{Leaves Section}}
At the innermost level of the recursion, the algorithm essentially converts coefficients modulo $P$ into much smaller coefficients, via Chinese remainder theorem.  We'll call the maximum (smaller) prime $q$.  Thus the conversion from values modulo $R \approx P^{1/4}$ to modulo a set of primes $q_k$ for varying $k$ is done only at the innermost portion of the recursion.  The essential idea is that the algorithm has a coefficient $c$ of a DFT as a number, which we can write as

\begin{align}
	c = x_0 R^0 + x_1 R^1 + \dots + x_{4L-1} R^{4L-1}
\end{align}

Using the same logic as Section \ref{Tabular Introduction Section}, the algorithm will use $8L$ tables, so that $x_k R^k$ is assigned a specific value for each of the smaller primes equal to or less than $q$.  In other words, $R^k \mod q_j$, for some small prime $q_j$, is one and only one value.  We can further assign $x_k R^k \mod q_j$ one and only one value.  Thus the algorithm can convert any coefficient into a system of primes via Chinese remainder theorem.  This requires no calculation during the actual algorithm, only calculation during the preprocessing.  Thus the time it takes is proportional to the size of the coefficient $c$, assuming constant time to access any bit of memory.


\Section{Algorithm Details\label{Algorithm Section}}
So, essentially we start with a multiplication algorithm that uses a number theoretic transform, or NTT, which is essentially a DFT in a finite field.  Our goal is to multiply two $v$-bit naturals, and we will assume that $v$ is approximately equal to $n \log{n} = \log{\nu}$, with more details to be described soon.  We used $v$ just to simplify the calculations with $n$.  One important thing to note is that our algorithm will need some precalculated information, and so we can perhaps describe it better as a circuit.

Our first real step towards finding some of the parameters that the algorithm will use is to determine the size of the largest of our small primes $q$, as noted above in Section \ref{Leaves Section}, which is used when we break the coefficients modulo $p$ into coefficients modulo smaller primes such as $q$.  We use $q$ as the size of the maximum (smaller) prime; that is, the primes that are used only in the innermost portion of the recursion.  To start, we'll assume that we can't use any tables larger than $n$ bits in total size, so that we are guaranteed not to take up too much time or space.  We now need to know how big the Chinese remainder theorem will allow us to make $q$.

We know that we can use the product of the smallest primes to compute a value modulo one large prime by using the Chinese remainder theorem.  Assuming that the smallest DFT that we use will have a size $m$, then we need to calculate a value of size $(p \cdot p)m$.  This is because each output coefficient takes an input that can range from $0$ to $p$, and multiplies it by another value, $\omega$, raised to some power, modulo $p$ again.  This gives us $p \cdot p$, which there are $m$ of these since we've assumed that the (small) DFT is of size $m$.  The \textit{primorial}, $Z\#$, is defined to be the product of the first $Z$ primes, and setting $Z\# = p^2 m$ and fining $\log_2{(Z\#)}$ will give us the bits in the largest of the smaller primes, $q$.  According to \cite{PrimorialBounds}, we have for $Z > 1$, as an upper bound on the primorial, that

\begin{align}
	\log{(Z\#)} &> Z \left( 1 - \frac{ 1 }{ 2 \log{Z} } \right) \\
							&> Z^{1 - o(1)}
\end{align}

This gives

\defeq{First Equation}
	Z^{1 - o(1)} > p^2 m
\eeq

We also know that (the log of) the sum of the first $x$ primes times $m$ gives us the bitsize of each entry of the tables, since we must have an entry for each prime times the number of entries, $m$.  We have a total of $\sum_{z=1}^Z{\text{Prime}(z)^m}$ total entries,\footnote{Here Prime$(z)$ means the $z$th prime.} since there are at most $z^m$ entries for each prime $z$ in the range from $1$ to $Z$.  We also know, from \cite{PrimeSum1}, that the sum of the first $Z$ primes is


\begin{align}
	\sum_{z=1}^Z{ \text{Prime}(z) } = \frac{Z^2}{2 \log{x}} + O \left( \frac{Z^2}{\log{Z}^2} \right)
\end{align}

Further, we have that for the sum of the $m$th powers of a prime, we have a formula from \cite{PrimeSumX}

\begin{align}
	\sum_{z=1}^Z{ \text{Prime}(z)^k } = \text{li}(Z^{k+1}) + O \left( Z^{k+1}e^{ -c \sqrt{ \log{Z} } } \right)
\end{align}

Here ``li'' is the \textit{offset} or \textit{Eulerian} logarithmic integral, as defined in \cite{website:Wikipedia-Logarithmic-Integral-Function}


\begin{align}
	\text{li}(Z) &= \int_2^Z{ \frac{du}{\log{u}} } \\
	             &= O \left( \frac{ Z }{ \log{Z} } \right)
\end{align}

Collecting functions, we have

\begin{align}
	\sum_{z=1}^Z{ \text{Prime}(z)^k } &= \text{li}(Z^{k+1}) + O \left( Z^{k+1}e^{ -c \sqrt{ \log{Z} } } \right) \\
	                                  &= O \left( \frac{ Z^{k+1} }{ \log{ Z^{k+1} } } \right) + O \left( Z^{k+1}e^{ -c \sqrt{ \log{Z} } } \right)
\end{align}

From Equation \ref{First Equation}


\begin{align}
	Z^{1 - o(1)} &> p^2 m \\
	Z            &> p^2 m
\end{align}

To get the size of all of our DFT tables, we set $n$ to be larger than the size of all of the tables, as demonstrated in Figure \ref{Equation Figure}.

\begin{PageFigure}
	\begin{center}
		\begin{align}
			n	&> \text{(Sum of Primes)($m$)(total table entries)} \\
				&> \left( \frac{Z^2}{2 \log{Z}} \right) (m) \left( O \left( \frac{ Z^{m+1} }{ \log{ Z^{m+1} } } \right) \right) \\
				&> \left( \frac{Z^2}{2 \log{Z}} \right) (m) c \left( \frac{ Z^{m+1} }{ \log{ Z^{m+1} } } \right) \\
				&> \frac{cmZ}{2} \left( \frac{ Z }{ \log{Z} } \right)^{m+2} \\
			2n/(mc) &> Z \left( \frac{ Z }{ \log{Z} } \right)^{m+2} \\
			2n/(mc) &> \left( \frac{ Z }{ \log{Z} } \right)^{m+3} \\
			(2n/(mc))^{1/(m+3)} &> \left( \frac{ Z }{ \log{Z} } \right) \\
			(2n/(mc))^{1/(m+3)} &> Z^{1-o(1)} \\
			(2n/(mc))^{1/m} &> Z^{1-o(1)} \\
			2n/(mc) &> Z^{m-o(1)} \\
			2n/c &> m Z^{m-o(1)} \\
			2n/c &> Z^{m-o(1)} \\
\label{Final-Equation}			2n/c &> (p^2 m)^{m-o(1)}
		\end{align}
	\end{center}
	\caption{\bf{Determining the Table Size}\label{Equation Figure}}
\end{PageFigure}





%

\begin{PageFigure}
	\begin{center}
		\defeq{$m$ Equation 1}
			\displaystyle \lim_{n \to \infty}{ \frac{2n/c}{ \left( \log{(n^{10} m)} \right)^m } } =
			\begin{cases}
				0      & \quad \text{if } m = \log{n} / \left( \left( \log{\log{n}} \right)^1 \right) \\
				\infty & \quad \text{if } m = \log{n} / \left( \left( \log{\log{n}} \right)^2 \right)
			\end{cases}
		\eeq
		\begin{align}
			\log{ \left( \frac{G n}{ \left( \log{(n^Hm)} \right)^m } \right) } &= \log{G} + \log{n} - m \left(\log{\log{(n^H m)}} \right) \\
			                                                                   &= \log{G} + \log{n} - \frac{ \log{n} }{ \left( \log{\log{n}} \right)^\alpha }\log{(H \log{n} + \log{m})} \\
																																				 &= \log{G} + \log{n} - \frac{ \log{n} }{ \left( \log{\log{n}} \right)^\alpha } \left( \log{H} + \log{\log{n}} + \log{ \left( 1 + \frac{\log{m}}{H \log{n}} \right) } \right)\\
		\end{align}
	\end{center}
	\caption{\bf{Determing $m$}\label{Limit Figure}}
\end{PageFigure}

From here, we can convert the inequality to a limit, taken from \cite{M.SE-Limit-Equation-1}:

\begin{align}
	2n/c > (p^2 m)^{m} \\
	\displaystyle \lim_{n \to \infty}{ \frac{2n/c}{ \left( \log{(n^{10} m)} \right)^m } }
\end{align}

This limit is calculated in Figure \ref{Limit Figure}, where we take the limit as $m$ approaches $\log{n}/(\log{\log{n}})^\alpha$.  When $\alpha > 1$, it is clear that the limit goes to infinity.  When $\alpha \le 1$, the subtracted term is greater than $\log{n}$ and the limit goes to $-\infty$.

\Subsection{Recursion Details\label{Recursion Section}}
For a DFT of size $\alpha$, we will use recursive calls of size $\sqrt{\alpha}$; more specifically, as $\lfloor \sqrt{\alpha} \rfloor$ or $\lceil \sqrt{\alpha} \rceil$.  This enables us to get as close to $\sqrt{\alpha}$ as possible, but at the expense of only being able to use certain DFT sizes.  Since we are always using $\Theta(\sqrt{\alpha})$ as the size of the inductive step of the recursion, we will make the original DFT of size $m^{2^s} \le n \le (m+1)^{2^s}$, where $m$ or $m+1$ is again the size of the base case DFT, and $n$ is the size of the original problem.  It should be fairly obvious that as $m$ grows larger, $m^{2^s}$ gets relatively closer to $(m+1)^{2^s}$.  It was already shown how $m$ grows as $n$ grows, for example, in Equation \ref{$m$ Equation 1}.

In addition to changing the size of $m$ as $n$ varies, we can also set $n = m^{2^s - x}(m+1)^Z$.  This just simply gives us more precise control of the size of $n$, since the ratio between similar sizes of $n$ is reduced to approximately $\label{Page Marker}(m+1)/m$.  


We next want to show how the speed of the base case affects the speed of the overall algorithm.  Ordinarily, a size $m$ FFT takes time $m \log{m}$, but our base case takes linear time, or $m \log{m} / \log{m}$.  For a size $m^2$ algorithm, there are approximately $2\sqrt{m^2} = 2m$ DFTs of size $m$.  So this takes time $2m \cdot m \log{m} / \log{m}$.  For $m^4$, this takes time

\begin{align}
	2\sqrt{m^4} \left( \cdot 2m \cdot m \log{m} / \log{m} \right) &= \\
	2m^2 \cdot 2m \cdot m \log{m} / \log{m} &= \\
	4m^4 \log{m} / \log{m} &= \\
	m^4 (4\log{m}) / \log{m} &= \\
	m^4 (\log{m^4}) / \log{m} 
\end{align}

In general, the algorithm takes time $n \log{n} / \log{m}$.  We've already shown this for the base cases $n = m^4$ and $n=m^2$.  So to prove this we use the induction.  Let the $n = m^{2^k}$ sized DFT (our algorithm) take time $m^{2^k} \log{m^{2^k}} / \log{m}$.  Then, we have for a size $n^2$ algorithm

\begin{align}
	2\sqrt{n^2} \text{(time for $n$ FFT)} &=\\
	2n (\cdot n \log{n} / \log{m}) &= \\
	2m^{2^k} \left(m^{2^k} \log{m^{2^k}} / \log{m} \right) &= \\
	2m^{2^{k+1}} \log{m^{2^k}} / \log{m} &= \\
	m^{2^{k+1}} 2\log{m^{2^k}} / \log{m} &= \\
	m^{2^{k+1}} \log{m^{2^{k+1}}} / \log{m}
\end{align}

\Subsection{Base Case Details}
The algorithm will proceed to use recursion with a base case of size $m$, and we have already shown how large $m$ can get before the table sizes are too large (i.e. bigger than $n$).  However, we want $n \approx m^{2^k}$ for some $k$ and $m$.  So first we find $\log_Z{n} \approx m$.  That is to say, we find the appropriate power of $Z$ such that $m^Z$ is approximately equal to $n$, our total transform size.  This is a simple, but time consuming process, so that we do it during preprocessing.  Then, we find $k$ such that $m^{2^k} \le n \le m^{2^{k+1}}$.  This is easy to find, since we pick $2^k \le Z \le 2^{k+1}$.

After finding the appropriate $k$, we do some fine-tuning.  We know that $m^{2^k} \le n$, so we pick a new $m_2$ so that ${m_2}^{2^k} \le n \le {m_2}^{2^{k+1}}$.  Finally, we use binary search to search for $z$ such that

\defeq{$m$ Equation}
	(m_2+1)^{2^k-z}{m_2}^z \le n \le (m_2+1)^{(2^k-z)+1}{m_2}^{z-1}
\eeq

This ensures that we're within a factor of $(m_2+1)/m_2$, as previously stated on page \pageref{Page Marker} in Section \ref{Recursion Section}.  Again, $m_2$ is slightly less than $m$, but is certainly greater than $\sqrt{m}$, since that is the size of the next level of recursion.\footnote{This is because at each level of the recursion, we set, as the recursion size, almost exactly the square root of the previous iteration size as the size of the next recursion.}  This ensures that the running time of the smallest DFT is not significantly affected by $m$.

\Subsection{Regarding $P$ and $N$}
Although we will use lookup tables of quartic roots of $v \approx n$, we need a larger prime $P$ to do our modular calculations.  To find out what $P$ is, we will use preprocessing.  However, to estimate the size of $P$, we make use of an analytic number theory theorem, called Linnik's theorem.


Essentially, we know that we want to use tables of size $v^{1/4} \times v^{1/4} \approx n^{1/2}$.  However, we need a prime that is at least as big as $n$.  We can start with \textit{Euler's totient function}, $\varphi{(n)}$.  It is well known that the \textit{multiplicative order} of any value modulo $n$ divides $\varphi{(n)}$.  The multiplicative order of a number is essentially the smallest power of a number, modulo some prime, that is equal to $1$.  We also know that $\varphi{(n)}$ divides $n-1$ for a prime $n$.  Putting all of this together, we want to find $P$ such that $n$ divides $P-1$.

To find this $P$, we can therefore examine the arithmetic progression $nd + 1$, for $n$ fixed and $d$ varying.  We want this number to be prime, in which case we will use it as $P$, and find a value modulo $P$ that has multiplicative order $n$, which we'll use as $\omega$.  This is where Linnik's theorem is required.  It states that any arithmetic progression where $n$ is fixed will have a prime $P$ such that $1 < P < O(n^L)$, where $L$ is Linnik's constant.  The current best bounds on $L$ is 5, \cite{Linnik's-constant-reference}, but if the Generalized Riemann hypothesis is true it is $\le \varphi{(n)}^2\log^2{n}$ \cite{Linnik's-constant-GRH}.


\begin{algorithm*}
	\begin{algorithmic}[1]
		\Procedure{PreprocessDFT}{ $n$ }
			\State Find $P$
			\State Find $\omega$
			\State Create Tables
		\EndProcedure
		
		\Procedure{DFT}{ $\{X\}$ }
			\State \Call{PreprocessDFT}{ $| \{X\} |$ }
			\State Recursively calculate DFTs
			\State \Return $\{ \widehat{X} \}$
		\EndProcedure
	\end{algorithmic}
	\caption{DFT Algorithm\label{DFT Algorithm}}
\end{algorithm*}

\Section{Runtime Analysis}
The major parts of the algorithm are shown in Algorithm \ref{DFT Algorithm}.

We start by analyzing the preprocessing.  To find a prime between $1$ and $n^L$, we can use the sieve of Atkin and Bernstein\cite{SieveOfAtkin}.  It takes $O(n^L)$ arithmetic operations to find all primes between $1$ and $n^L$.

Next, we find an $\omega$ that has multiplicative order $P \approx n^L$.  This takes at most time $O(n^L)$, according to the following algorithm.  First, we build a linked list of all numbers.  Then we pick a number at random.  We then repeatedly find powers of that number, erasing their presence from the list.  If we find a number with multiplicative order $n$ when we arrive at $1$, we are done.  Otherwise, we continue and pick another number.  Hopefully we arrive at one after $n$ iterations.  If not, we either continue as long as we haven't yet eliminated $n$ numbers, or if we have eliminated more than $n$, we need to do more.  In this case, we take the total multiplicative order of our base number we used, call it $n_2$.  We take the $n_2 / n$th power of this number, and this will then be $\omega$.  This all takes time $O(n^L)$.

However, if we guess at values for $\omega$, our $P$th root of unity, we can find an omega in randomized time $O(\sqrt{P} / \varphi{(P-1)})$, and even in time $O(\sqrt{R} / \varphi{(P-1)})$, where $F$ is the largest factor of $\varphi{(P)} = P-1$, where $\varphi{(P)}$ is Euler's totient function.  The reason is that the discrete logarithm runs in time $O(\sqrt{P})$ for many different algorithms, and it is well known that there are $\varphi{(P-1)}$ elements of maximum order modulo $P$.  According to \cite{website:Wikipedia-Pollard-rho}, when combined with the Pohlig-Hellman algorithm, the running time for the discrete logarithm is $O(\sqrt{F})$, thus the second, faster runtime.


The table creation time is $O( 8L\sqrt{n})$ arithmetic operations for the multiplication tables for the usual DFT multiplications.  This is because there are $4L$ different inputs of size $n$, times $2$ because the multiplication operation at most doubles the input words to yield $8L$ output words.

We know the table creation time for the small prime at the leaves of the recursion is proportional to $n \log{m}$, since we've already established that the size of all of the tables to convert between $P$ and the small primes $q_k$ is less than $n$.  This also follows because all of the DFTs that are used are size $m$, as was discussed in Section \ref{Algorithm Section}, and thus this will multiply the size of the tables by at most $\log{m}$.

Thus, summing all of the preprocessing times together, we come up with an $O(n^L)$ time for the preprocessing.

\Subsection{Algorithm Speed}


We already have explored much of the time required to compute the DFT.  We use recursion to reach a DFT of size $m$, and that DFT takes $O(m)$ operations, times at most $O(\log{n})$.  So the total time for the new DFT algorithm takes time $O(m \log{n})$.  Now, according to Section \ref{Recursion Section}, this makes the total time be at most $O(n \log{n} / \log{m})$ arithmetic operations.  This is almost the speed of the algorithm.  Recall that we set $n \log{n} \approx v$.  Thus we have $v / \log{m}$ operations, which each take time $O(\log{P})$, so that we have $v \log{v} / \log{m}$ time, and this is the speed of the algorithm.

That is to say, with our calculations for $m$ from Equation \ref{$m$ Equation 1} in Section \ref{Algorithm Section} on page \pageref{$m$ Equation}, the time is

\begin{align}
	\rho \log{\rho} / \log{m} &< \\
	\rho \log{\rho} / \left( \log{  \frac{ \log{2n/c} }{ \text{N}\left( p^2 \log{2n/c} \right) }  } \right) &< \\
	\rho \log{\rho} / \left( \log{  \frac{ \log{2v/c} }{ \text{N}\left( {v^L}^2 \log{2v/c} \right) }  } \right) &= \\
	\rho \log{\rho} / \left( \log{  \frac{ \log{2v/c} }{ \text{N}\left( {v^5}^2 \log{2v/c} \right) }  } \right)
\end{align}

Using Mathematica again shows that:

\defeq{Mathematica Limit}
	\lim_{v \to \infty} \frac{  v \log{v}  }{  \log{\frac{ \log{2v/c} }{ \text{R}\left( {v^5}^2 \log{2v/c} \right)}}  } = 0
\eeq

So we are assured that this bound is better than the previous $O(v \log{v})$ bound.  This makes the time be $O(v \log{v} / \log{m}$, which equals $O(v \log{v} / (\log{\log{v}} / \log{\log{\log{v}}}))$


\appendix

\bibliographystyle{plain}
\bibliography{BibFile07000000027}



\end{document}